\documentclass[twocolumn,showpacs,preprintnumbers,amsmath,amssymb]{revtex4}

\usepackage{graphicx}
\usepackage{dcolumn}
\usepackage{bm}

\begin{document}

\title{Strongly coupled large-angle stimulated Raman scattering of short
laser pulse in plasma-filled capillary}
\author{Serguei Kalmykov}
\altaffiliation{Present address: Department of Physics and
Institute for Fusion Studies, The University of Texas at Austin,
One University Station C1500, Austin, Texas 78712; e-mail:
kalmykov@physics.utexas.edu} \affiliation{Centre de Physique
Th\'eorique (UMR 7644 du CNRS), Ecole Polytechnique, 91128
Palaiseau cedex, France, } \affiliation{Max-Planck-Institut f\"ur
Quantenoptik, D-85748 Garching, Germany}
\author{ Patrick Mora}
\affiliation{Centre de Physique Th\'eorique (UMR 7644 du CNRS),
Ecole Polytechnique, 91128 Palaiseau cedex, France }
\date{\today}

\begin{abstract}
Strongly coupled large-angle stimulated Raman scattering (LA SRS)
of a short intense laser pulse develops in a plane plasma-filled
capillary differently than in a plasma with open boundaries.
Coupling the laser pulse to a capillary seeds the LA SRS in the
forward direction (scattering angle smaller than $\pi/2$) and can
thus produce a high instability level in the vicinity of the
entrance plane. In addition, oblique mirror reflections off
capillary walls partly suppress the lateral convection of
scattered radiation and increase the growth rate of the SRS under
arbitrary (not too small) angle. Hence, the saturated convective
gain falls with an angle much slower than in an unbounded plasma
and even for the near-forward SRS can be close to that of the
direct backscatter. At a large distance, the LA SRS evolution in
the interior of the capillary is dominated by
quasi-one-dimensional leaky modes, whose damping is related to the
leakage of scattered radiation through the walls.
\end{abstract}
 \maketitle
\newpage

\section{\label{Sec1}Introduction}
The technique of chirped-pulse amplification~\cite{Mourou} made
sub-picosecond laser pulses of high power ($P>10^{12}$ W)
available for generation of coherent x-rays~\cite{Burnett}, high
harmonics of radiation~\cite{Lompre}, and laser wakefield
acceleration (LWFA) of electrons~\cite{Tajima,Mora,Esarey_IEEE} in
rarefied plasmas [where $\omega_0\gg\omega_{pe}$, $\omega_0$ is a
laser frequency, $\omega_{pe}=(4\pi e^2n_0/m_e)^{1/2}$ is an
electron plasma frequency, $n_0$ is a background electron density,
$m_e$ and $-|e|$ are the electron mass at rest and charge]. Full
potential of these applications can be realized with the
laser-plasma interaction length increased beyond the Rayleigh
diffraction length, $z_R= \pi \sigma_0^2/\lambda_0$,  by means of
external optical guiding~\cite{Esarey_IEEE} (here and later,
$\lambda_0\approx 2\pi c/\omega_0$ is a laser wavelength, and
$\sigma_0$ is a laser beam waist radius). In particular, a
dielectric capillary, where the oblique mirror reflections
suppress the laser beam diffraction, can be used as a guiding
tool~\cite{Marcatili,Andreev,Capillary_vacuum}. Plasma can be
created in a capillary by an optical field ionization of the
filling gas~\cite{Capillary_plasma,CrosPScr,Andreev_capillary}, or
by a laser ablation of the walls~\cite{Kitagawa}. Then, the
large-angle stimulated Raman scattering (LA SRS) starts to
challenge the transportation of a laser beam over a long
distance~\cite{McKinstrie3,CourtoisDiss}.

In the standard SRS process~\cite{Andreev0}, the pump
electromagnetic wave (EMW) is scattered off spontaneous
fluctuations of electron density, which, in turn, can be amplified
by the ponderomotive beat wave of pump and scattered light.
Appropriate phase matching of the waves results in a positive
feedback loop with the onset of a spatio-temporal
instability~\cite{Forslund}. When the plasma extent is much larger
than a laser pulse length, and no reflections off plasma
boundaries occur, both scattering electron plasma waves (EPW) and
scattered EMW quit the region of amplification, and the convective
gain saturates within a time interval of the order of pulse
duration~\cite{Mora0,McKinstrie1,Kalmykov_multi,Kalmykov}. Given
the pulse length, the maximum possible gain remains the same for
{\it all} scattering angles, and whether it is achieved or not for
a given angle is determined by the laser pulse aspect ratio
only~\cite{Mora0,McKinstrie1,Kalmykov_multi,Kalmykov}. Convection
of scattered radiation out of the laser waist may result in a
strong pulse depletion~\cite{Rousseaux}. Even when the full
depletion does not occur, the LA SRS can produce considerable
pulse erosion~\cite{ErrDep}, suppression of the relativistic
self-focusing~\cite{Tzeng}, heating and pre-acceleration of plasma
electrons~\cite{Kruer}, and seeding the forward
SRS~\cite{Sakharov,Hafizi}. Thereby, knowing the details of the LA
SRS evolution in various physical conditions is a matter of high
importance for applications.

Confining plasma between reflecting surfaces deeply modifies the
SRS process. In the one-dimensional (1D) geometry, the Raman
backscatter changes its nature from convective to
absolute~\cite{Bobroff}: reflections trap the unstable radiation
modes inside plasma and give rise to the continuous amplification.
When the laser beam is confined between the mirror-reflecting
partly transparent walls, and propagates collinearly to them,
reflections reduce the sideward convection of scattered light. If
the reflective modes dominate in plasma, the LA SRS gain tends to
that of the direct backscatter and thus reveals a dramatic
increase in comparison with the open-boundary
system~\cite{McKinstrie3}. The LA SRS in this geometry has been
considered so far in the regime of weak
coupling~\cite{McKinstrie3}, when the scattering EPW is similar to
the plasma natural mode~\cite{Forslund}, and temporal growth rate
is well below the electron plasma frequency. This regime requires
fairly low amplitude of a laser pulse, i.e., $a_0\ll
\sqrt{\omega_{pe}/\omega_0}\ll1$ [$a_0=eE_0/(m_e\omega_0c)$ is a
normalized amplitude of the laser electric field]. However, for
the efficient LWFA~\cite{Tajima,Mora,Esarey_IEEE}, the plasma
density have to be reduced in order to increase the gamma-factor
of the laser wakefield, $\gamma_g=\omega_0/\omega_p$, and, hence,
the electron energy gain. With $1/\gamma_g<a_0^2<1$, the LA SRS
becomes strongly coupled: its temporal growth rate exceeds
$\omega_{pe}$, and the scattering EPW differs from the natural
mode of plasma oscillations
~\cite{Mora0,Kalmykov,Sakharov,Darrow,MounaixSC}. In a
plasma-filled capillary, the strongly coupled LA SRS  acquires new
specific features: in a wide range of parameters relevant to the
self-modulated LWFA~\cite{Mora}, our PIC simulations (using the
code WAKE~\cite{Mora1}) discovered a vast enhancement of the
near-forward SRS in the immediate vicinity of the capillary
entrance aperture. The unstable plasma modes were primarily
transverse and therefore useless for the longitudinal electron
acceleration. For the same range of parameters, this effect has
never been significant in an open-boundary plasma. Independent
fluid modelling~\cite{AndreevPrComm} verified these observations.

Making a step to understanding this phenomenon we propose a
two-dimensional (2D) linear theory of strongly coupled SRS of a
short laser pulse under a given angle $\alpha$ in a slab of
rarefied plasma laterally confined between the mirror-reflecting
partly transparent flat walls (flat capillary). Boundary
conditions for the scattered radiation describe the oblique mirror
reflections and the electromagnetic (EM) seed at the entrance
plane. We associate the latter with the signal formed of the
high-order capillary eigenmodes produced by the laser beam
coupling to the capillary (the coupling process is described
elsewhere~\cite{Andreev,CrosPScr} and is outlined in
Appendix~\ref{App1}). SRS in forward ($\alpha<\pi/2$) and backward
($\alpha>\pi/2$) direction proceed differently. Forward Raman
amplification of the EM seed can be dominant within a finite
distance from the entrance plane and be responsible for the
instability enhancement observed in the modelling. On the other
hand, the backward SRS is affected by the reflections only. As the
LA SRS of a finite-length laser pulse preserves the convective
nature (backward and, partly, sideward convection of radiation is
allowed), the gain saturation occurs within a finite distance from
the entrance plane. Reflections give the unstable modes additional
rise time in any transverse cross-section of a plasma, and, even
for relatively small scattering angles (such as $\alpha=\pi/6$
taken for numerical examples of this paper), the saturated
convective gain can approach that of the backward SRS (BSRS). The
field structure is then approximated by a quasi-1D lossy mode
whose damping is produced by the leakage of radiation through the
walls.

The paper is organized as follows. Section~\ref{Sec2} presents a
theoretical model of the strongly coupled LA SRS in a  2D slab
geometry. The laser pulse entrance into a plasma and oblique
mirror reflections of scattered light are expressed in terms of
appropriate boundary-value conditions for the coupled-mode
equations. General solution of the boundary-value problem is
presented (derivation is given in Appendix~\ref{App2}).
Section~\ref{Sec3} discusses the spatio-temporal evolution of
instability. The case of fully transparent lateral boundaries is
considered in subsection~\ref{SubSec3.1}. Enhancement of the LA
SRS in the generic reflective case is considered, and the maximum
gain factors are evaluated in terms of appropriate asymptotic
solutions in subsection~\ref{SubSec3.2}. Section~\ref{Sec4}
summarizes the results.

\section{\label{Sec2}Basic equations and solution of boundary-value problem}

\begin{figure}
\includegraphics[scale=1]{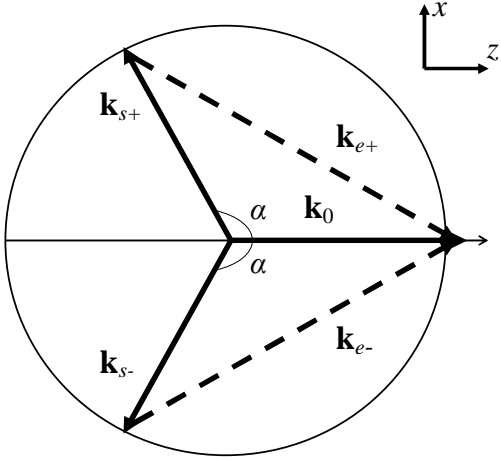}
\caption{\label{Fig1} Wave vector diagram of the LA SRS;
``+''(``-'') stands for the up-(down-)going interaction.}
\end{figure}

In a 2D plasma slab confined between mirror-reflecting capillary
walls, the high-frequency (hf) electric field is a superposition
\begin{eqnarray}
{\bf a}({\bf r},t)  & = & \frac{\displaystyle e^{-i\omega_0t}
}{2}\left\{{\bf a}_0({\bf r},t)e^{ik_0z}  +
\sum\limits_{\sigma=\pm}{\bf a}_{s\sigma}({\bf r},t)e^{i({\bf
k}_{s\sigma},{\bf r})} \right\}\nonumber\\&  + & c. c.\label{1}
\end{eqnarray}
of that of the laser, ${\bf a}_0$, and of up- and down-going
scattered EMWs, ${\bf a}_{s\pm}\equiv e{\bf
E}_{s\pm}/(m_e\omega_0c)$ ($|{\bf a}_{s\pm}|\ll |{\bf a}_{0}|<1$);
in Eq.~(\ref{1}) and below, ${\bf r}=(x,z)$ is a radius-vector in
a plane geometry. We assume that the waves~(\ref{1}) have the
linear polarization with the electric vectors parallel to the
walls, and that the scattering occurs under an angle $\alpha$ in
the plane orthogonal to the polarization vector. In the rarefied
plasma, both ${\bf k}_0={\bf e}_z k_0$ and ${\bf k}_{s\pm}$ obey
the same dispersion relation
$\omega_0^2=\omega_{pe}^2+c^2k_{0(s\pm)}^2\approx c^2
k_{0(s\pm)}^2$; hence, $|{\bf k}_{s\pm}|\equiv k_s =k_0$,
${k_{s\pm}}_z\equiv {k_s}_z=k_0\cos\alpha$, $\pm{k_{s\pm}}_x\equiv
{k_s}_x=k_0\sin\alpha$, and  the amplitudes $a_{0(s\pm)}$ vary
slowly in time and space on the scales $\omega_0^{-1}$,
${k_s}_z^{-1}$, and ${k_s}_x^{-1}$. Ions form a homogeneous
positive background; this assumption holds for a laser pulse
shorter than an ion plasma period, $t_0\ll2\pi\omega_{pi}^{-1}$.
The beat wave of incident and scattered radiation excites
perturbations of electron density,
\begin{equation}
\label{2A} \frac{n_e-n_0}{n_0}= \sum\limits_{\sigma=\pm}
N_\sigma^*({\bf r},t)e^{i({\bf k}_{e\sigma},{\bf r})} + c.c.,
\end{equation}
whose wave vectors obey the matching conditions ${\bf
k}_{e\pm}={\bf k}_0-{\bf k}_{s\pm}$; hence, $|{\bf
k}_{e\pm}|\equiv k_e =2k_0\sin(\alpha/2)$, ${
k_{e\pm}}_z\equiv{k_e}_z=k_0(1-\cos\alpha)$, ${ k_{e\pm}}_x=\mp
k_0\sin\alpha$. Wave vector diagram of the LA SRS is shown in
Fig.~\ref{Fig1}. The scattering EPW has a longitudinal component
of phase velocity small compared to the speed of light, i.e.,
${k_e}_z > k_p\equiv\omega_{pe}/c$. This restriction eliminates
the forward Raman scattering~\cite{Sakharov,Turano} and resonant
modulational instability (RMI)~\cite{RMI}. In the rarefied plasma,
the amplitudes $N_\pm$ vary slowly in space on the scales
${k_e}_{z}^{-1}$, ${k_e}_{x}^{-1}$.

The amplitudes of up- and down-going scattered EMW and scattering
EPW obey the coupled-mode equations derived from the equations of
non-relativistic hydrodynamics of cold electron fluid in the hf
field~(\ref{1}) and Maxwell's equations for the scattered
radiation,
\begin{subequations}
\label{A}
\begin{eqnarray}
 i\left(\frac{\partial}{\partial \xi}
+V_z\frac{\partial}{\partial z} \pm V_x\frac{\partial}{\partial
x}\right) a_{s\pm} & = & g_1 N_\pm,
\label{3}\\
-\left( \frac{\partial^2}{\partial \xi^2}+k_p^2\right)N_\pm & = &
g_2a_{s\pm},\label{4}
\end{eqnarray}
\end{subequations}
where $V_z=\cos\alpha/(1-\cos\alpha)$,
$V_x=\sin\alpha/(1-\cos\alpha)$, $g_1=(a_0/2)(k_p^2/{k_e}_z)$, and
$g_2= a_0^*(k_e/2)^2$. The wave coupling parameter is $G^3\equiv
g_1g_2=(a_0/2)^2k_p^2k_0$  (strong coupling is the case for $G\gg
k_p$). Equations~(\ref{A}) are expressed through the variables
$x$, $z$, and $\xi=ct-z$, that is, the temporal evolution of waves
is traced in an $x-y$ cross-section at a longitudinal position
$z$.

\begin{figure}
\includegraphics[scale=0.9]{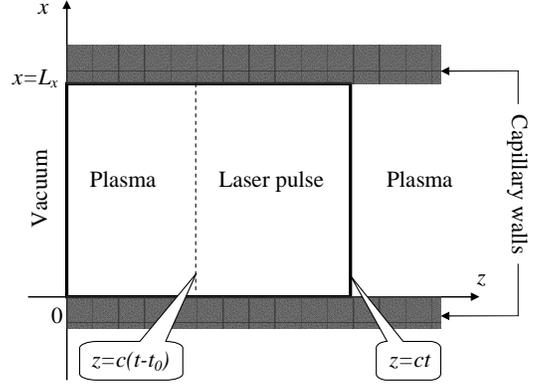}
\caption{\label{Fig2} Geometry of a laser pulse propagation in a
laterally confined plasma. Laser pulse enters the plasma at $z=0$
and $t=0$, and moves towards positive $z$. Boundary conditions are
posed at the plasma boundary $z=0$, the pulse leading front $z=ct$
and the walls $x=0$ and $x=L_x$. Rear edge of the pulse
$z=c(t-t_0)$ is a free boundary through which the waves quit the
region of amplification. The boundary-value problem is solved in
the area $c(t-t_0)<z<ct$, $0<x<L_x$. }
\end{figure}

Figure~\ref{Fig2} shows the interaction area.  At $z=0$, $t=0$,
the laser pulse enters a semi-infinite plasma-filled gap between
flat mirror-reflecting walls ($z\ge0$, $0 \le x\le L_x$) and
propagates towards positive $z$. The pulse leading front, $\xi=0$,
encounters the stationary level of electron density perturbations
with a constant amplitude $N_0$,
\begin{subequations}
\label{B1}
\begin{eqnarray}
N_\pm (x,z,0)& = & N_0,  \label{9}\\
\partial N_\pm/\partial\xi (x,z,0) & = & 0,\label{10}
\end{eqnarray}
\end{subequations}
fluctuations of radiation in fresh plasma being neglected,
\begin{equation}
a_{s\pm}(x,z,0) \equiv0. \label{B2}
\end{equation}
At the capillary entrance plane, the transverse profile of
radiation can have a significant content of the high-order
capillary eigenmodes (coupling the incident laser beam to the
capillary is discussed elsewhere~\cite{Andreev,CrosPScr} and is
outlined in Appendix~\ref{App1}). The resonant condition for the
wave vectors selects the modes that can be amplified by the
forward  SRS ($\alpha<\pi/2$) thus leading to the formation of an
EM seed signal. To facilitate the forthcoming analytic job, we
give this signal in a simple parabolic form with an amplitude
vanishing at the walls in order to provide the continuity of the
solution in the interior of the capillary,
\begin{equation}
a_{s\pm}(x,z=-0,\xi) = a_{s0}\left[1-(1-2x/L_x)^2\right].
\label{B21}
\end{equation}
Inside the capillary, oblique mirror reflections  couple up- and
down-going EMW: each reflection converts an up-going wave into a
down-going one and vice versa,\begin{subequations} \label{B3}
\begin{eqnarray}
a_{s+}(0,z,\xi) & = & r(\alpha)\,a_{s-}(0,z,\xi), \label{7}\\
a_{s-}(L_x,z,\xi) & = & r(\alpha)\,a_{s+}(L_x,z,\xi). \label{8}
\end{eqnarray}
\end{subequations}
The conditions~(\ref{B3}) set up a quasi-1D exponential behavior
of waves at large $z$. The reflectivity coefficient is a known
function of scattering angle,
$r(\alpha)=|\sin\alpha-[(\delta_w/\delta_{pl})^2-
\cos^2\alpha]^{1/2}|/\{\sin\alpha+[(\delta_w/\delta_{pl})^2-
\cos^2\alpha]^{1/2}\} $, where $\delta_w$ and $\delta_{pl}$ are
the refraction indexes of walls and plasma. Figure~\ref{Fig3}
shows $r(\alpha)$ for a glass capillary with $\delta_w\approx1.5$
and $\delta_{pl}\approx1$.

The temporal increment of strongly coupled LA SRS exceeds the
electron plasma frequency. Hence, we neglect $k_p^2 N_\pm$ in
comparison with $\partial^2 N_\pm/\partial \xi^2$ in the left-hand
side (LHS) of Eq.~(\ref{4}). With the allowance for not very tight
capillary, the pump field envelope $a_0(x,\xi)$ represents a
portion of laser radiation coupled to the capillary which
experiences mostly paraxial propagation,
$k_\perp/k_0\ll\sqrt{\omega_{pe}/\omega_0}$. Assuming that the
pump field evolution at a given point $(x,z)$ takes much longer
than the SRS growth ($z_R/c\gg t_0$),  and in order to enable the
analytic progress, we approximate $a_0(x,\xi)$ with a fixed flat
profile at any position $z$ in a capillary of the width $L_x$,
$a_0(x,\xi)=a_0H(x)H(L_x-x)H(\xi)H(ct_0-\xi)$, using the effective
pulse duration $t_0$ and amplitude $a_0$. Here and below, $H(y)$
is the Heaviside step-function. Solution of the boundary-value
problem,\begin{widetext}
\begin{subequations} \label{C}
\begin{eqnarray}
a_{s+}({\bf R};r)  &  =  & - a_{s0} \mathcal{F}({\bf R};r)
{\vphantom1}_0\tilde{F}_2(;1,1/2;i\zeta) - iN_0 \frac{g_1}{3}
\sum\limits_{j=1}^3 \frac{e^{c_j\xi}}{c_j} [1- \Phi_{1D}({\bf
R};r,c_j)-\Phi_{2D}({\bf
R};r,c_j)]  ,\label{11}\\
N_+({\bf R};r) &  =  & \frac14a_{s0}g_2\mathcal{F}({\bf R};r)(\xi
- z/V_z)^2 {\vphantom1}_0\tilde{F}_2(;2,3/2;i\zeta) +
\frac{N_0}{3}\sum\limits_{j=1}^3 e^{c_j\xi}[1  -  \Phi_{1D}({\bf
R};r,c_j)  -  \Phi_{2D}({\bf R};r,c_j)],\label{12}
\end{eqnarray}
\end{subequations}
is then obtained via the 2D Laplace transform; here, ${\bf
R}=({\bf r},\xi)$, $c_j^3=iG^3$,
$\zeta=(G^3/4)(\xi-z/V_z)^2z/V_z$,
${\vphantom1}_0\tilde{F}_2(;b_1,b_2;\theta)$ is the regularized
generalized hypergeometric function~\cite{IncGamma},
\begin{subequations}
\label{D}
\begin{eqnarray}
\Phi_{1D} & = & F_s\left(c_j,\frac{z}{V_z},\xi\right)
\left\{H(V_zx \!- \!V_xz) \!+\!  \sum\limits_{n=1}^{\infty}
r^n [H(V_xz\! - \!V_zx_{n-1}) \! - \! H(V_xz \! - \! V_zx_n)] \right\},\label{13}\\
\Phi_{2D} & =& (1-r)\sum\limits_{n=0}^{\infty}r^n
F_s\left(c_j,\frac{x_n}{V_x},\xi\right)H(V_xz-V_zx_n),\label{15}\\
\mathcal{F} & = &
\sqrt{\pi}H\left(\xi-\frac{z}{V_z}\right)\Biggl\{\frac{(V_zx-
V_xz)(V_zx-V_xz-V_zL_x)}{(V_zL_x/2)^2}H(V_zx-V_xz)\nonumber\\
&+& \sum\limits_{n=1}^{\infty} r^n
\frac{(V_zx_{n-1}\!-\!V_xz)(V_zx_n\!-\!V_xz)}{(V_zL_x/2)^2}\left[H(V_x
z\!-\!V_zx_{n-1})\!-\!H(V_xz\!-\!V_zx_n)\right]\Biggr\},\label{14}
\end{eqnarray}
\end{subequations}\end{widetext}
where  $x_n=x+nL_x$, and the fundamental solution $F_s(\mu,v,\xi)$
(Ref.~\onlinecite{Kalmykov}) is defined by Eq.~(\ref{A_11}). The
up- and down-going amplitudes are symmetric, $a_{s-}(x) =
a_{s+}(L_x - x)$, $N_-(x)=N_+(L_x-x)$, so we consider below the
evolution up-going waves only. Our solution possesses the same
generic structure as the weakly coupled reflective solution
discussed in detail in Ref.~\onlinecite{McKinstrie3}. However,
contrary to the case of semi-infinite laser pulse of
Ref.~\onlinecite{McKinstrie3}, the growth time of unstable waves
is now limited by the pulse duration $t_0$, and, as shown in the
subsection~\ref{SubSec3.2}, the gain at a given point $(x,z)$
remains finite and is given by either Eq.~(\ref{B12})
or~(\ref{19}).

\section{\label{Sec3} Spatio-temporal evolution of unstable waves}

\subsection{\label{SubSec3.1}Evolution of instability in open-boundary system}

When the lateral boundaries are fully transparent, $r=0$, the EM
seed at the entrance plane vanishes ($a_{s0}=0$), and the
instability grows from the electron density noise in a fresh
plasma ahead of the pulse. The functions~(\ref{D}) then read
\begin{eqnarray*}
\Phi_{1D}({\bf R};r,c_j) & = & F_s(c_j,z/V_z,\xi)H(V_zx-V_xz),\\
\Phi_{2D}({\bf R};r,c_j) & = & F_s(c_j,x/V_x, \xi)H(V_xz-V_zx).
\end{eqnarray*}
Plasma is divided by the characteristics $\xi=z/V_z$,
$x=z(V_x/V_z)$, $\xi=x/V_x$, into ranges of dependence, where the
solution is prescribed by the boundary conditions for radiation
posed at the pulse leading edge $\xi=0$ (range~{\bf I}),
\begin{equation}
\label{11B}\left.
\begin{array}{rcl}
\xi & < & x/V_x\\
\xi & < & z/V_z
\end{array}\right\}
 \Longrightarrow  \Phi_{1D}\equiv0,\,\,\Phi_{2D} \equiv0
\end{equation}
at the wall $x=0$ (range~{\bf II}),
\begin{equation}
\label{12B}\left.
\begin{array}{rcc}
x & < & z(V_x/V_z)\\
\xi & > & x/V_x
\end{array}\right\}
 \Longrightarrow   \Phi_{1D}\equiv0,\,\,\Phi_{2D}
 \not\equiv0
 \end{equation}
or at the entrance plane $z=-0$ (range~{\bf III}),
\begin{equation}
\label{13B}\left.
\begin{array}{rcc}
x & > & z(V_x/V_z)\\
\xi & > & z/V_z
\end{array}\right\}
  \Longrightarrow  \Phi_{1D}\not\equiv0,\,\,\Phi_{2D}
 \equiv0.
 \end{equation}
The principal feature that makes the LA SRS of a short
pulse~\cite{Kalmykov} different from the case of semi-infinite
laser beam~\cite{McKinstrie1} is the gain saturation within a
finite distance from the entrance plane. At some point,
$z_{sat}<+\infty$, the scattered radiation arriving from the
plasma boundary $z=-0$ drops behind the laser pulse, and in all
the points $z>z_{sat}$ neither part of the pulse belongs to the
range {\bf III}. Then, the evolution of waves and the gain do not
alter with $z$. The gain saturates differently for the forward
($\alpha<\pi/2$) and backward ($\alpha>\pi/2$) scattering.

\begin{figure}
\includegraphics[scale=1]{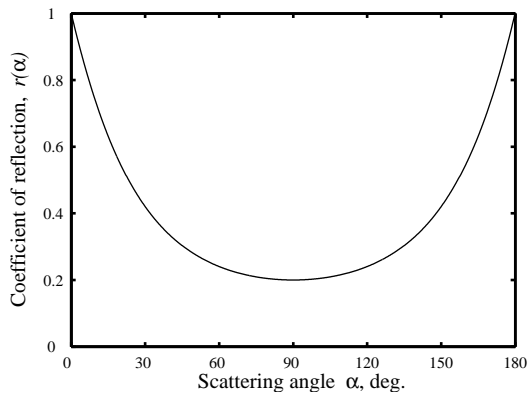}
\caption{\label{Fig3} Coefficient of reflection versus scattering
angle for a glass wall with an index of refraction $\delta_w=1.5$
(the refraction index of plasma is taken equal to unity).}
\end{figure}

When $\alpha<\pi/2$, and the distance from the entrance plane is
not too large, i.e., $z<\min\left\{V_zct_0,L_x(V_z/V_x)\right\}$,
all the three areas~(\ref{11B})-(\ref{13B}) are available in the
pulse body (that is, within a rectangle $0\le\xi\le ct_0$, $0\le
x\le L_x$). Given the point $(x,z)$, waves fall initially within a
range~{\bf I}, where they grow in time exponentially with an
angle-independent increment
\begin{equation}
\label{13D} \gamma_0=(\sqrt{3}c/2)G\approx(\sqrt{3}/2)
\sqrt[3]{(a_0/2)^2\omega_0\omega_{pe}^2}.
\end{equation}
Note, that $\kappa=\gamma_0/c$ is the known ``spatial'' increment
of the strongly coupled BSRS in the co-moving
frame~\cite{Esarey_IEEE,Mora0,Kalmykov_multi,Kalmykov,Sakharov,MounaixSC}.
The evolution of waves is strictly 1D in space on this stage.
Later, information from the boundaries $x=0$ and $z=-0$ reaches
the point $(x,z)$, and the spatial dependence becomes either 2D
for $\xi>x/V_x$, $x<z(V_x/V_z)$ (range {\bf II}) or remains 1D for
$\xi> z/V_z$, $x>z(V_x/V_z)$ (range {\bf III}). The waves are not
exponentially growing at this time. In the range {\bf III}, the
entrance effect dominates: vanishing the scattered EMW at the
boundary $z=-0$ determines the behavior of 1D amplitudes. Deeply
enough in plasma, $z\ge \min\left\{V_zct_0,L_x(V_z/V_x)\right\}$,
the entrance effect vanishes as the pulse terminates sooner than
the scattered EMW from the entrance plane can reach the observer
at a given $z$. The pulse body is then divided between the ranges
{\bf I} and {\bf II}, and the evolution of LA SRS is the same
through the rest of the plasma.

For $\alpha>\pi/2$, the boundary-value condition posed for
radiation at $z=-0$ can only produce the scattered EMW convecting
outwards (the EMW characteristic $\xi=z/V_z$ recasts in the lab
frame variables as $z=-ct|\cos\alpha|$, and corresponds to the
wave propagating towards negative $z$). Thus, the entrance effect
does not change the solution at a positive $z$, and the
boundary-value condition~(\ref{B21}) becomes excessive (this is
also valid in the reflective case). The spatio-temporal evolution
of the LA SRS remains the same at any $z\ge0$.

Given the pulse duration, the maximum of the saturated gain,
$a_{s+},N_+\sim e^{\gamma_0t_0}$,  does not depend on the
scattering angle, and whether it is achieved or not is determined
solely by the pulse aspect ratio~\cite{Kalmykov}. If the pulse is
wide, or the scattering angle is sufficiently large,
$\alpha>\alpha_0 =2\arctan\left(ct_0/L_x\right)$, the maximum gain
is achieved at the pulse rear edge $\xi=ct_0$ for
$ct_0\cot(\alpha/2)<x<L_x$. Hence, the angular spectrum of
scattered light is prescribed by the pulse aspect ratio rather
than the angular dependence of the increment. For $L_x\gg ct_0$,
scattering within a broad range of angles $2ct_0/L_x<\alpha\le\pi$
proceeds with the maximum gain. Otherwise, for $L_x\ll ct_0$, the
highest gain corresponds to the near-backward scattering only,
$\pi-L_x/(ct_0)<\alpha\le\pi$ (Ref.~\onlinecite{Kalmykov}). To
estimate the pulse energy depletion due to the LA SRS, it is
sufficient to neglect the radiation scattered under angles smaller
than $\alpha_0$ (Ref.~\onlinecite{Kalmykov_d}).

\subsection{\label{SubSec3.2}LA SRS evolution in the reflective case}

In a capillary, the oblique mirror reflections of scattered light
off the walls [the boundary condition~(\ref{B3})] contribute to
the LA SRS evolution over the whole laser path in plasma, but
become actually dominating later, when the entrance effect
vanishes ($z>V_zct_0$). The reflections establish a long-distance
asymptotic state of the LA SRS --- amplification of a quasi-1D
radiation and plasma modes with a temporal increment close to that
of the direct backscatter~(\ref{13D}). Besides, the forward SRS is
seeded by the EM signal $a_{s+}(x,-0,\xi)$ at the capillary
entrance plane [the boundary condition~(\ref{B21})]. If the signal
amplitude $a_{s0}$ exceeds the amplitude of the electron density
noise $N_0$, the unstable waves will achieve a large amplitude (by
virtue of the high seed level) at the transient stage of laser
propagation, $0<z<V_zct_0$. In this case, contrary to the SRS in
the unbounded plasmas described in the preceding subsection, the
near-forward SRS exhibits much higher level of amplification than
the backward SRS ($\alpha>\pi/2$), and its contribution to the
dynamics of electron density perturbations can be dominating.
Nonlinearities of plasma response that can then appear are worth
investigating and will be addressed in future publications.

\begin{figure}
\includegraphics[scale=1]{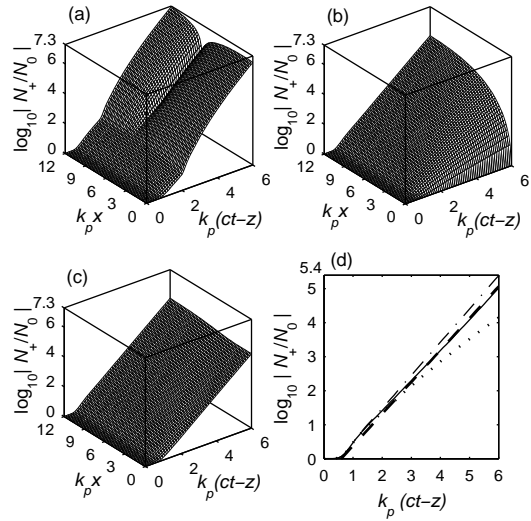}
\caption{\label{Fig4} Spatio-temporal evolution of the up-going
EPW in the field of transversely limited laser pulse of finite
duration; the pulse aspect ratio is $ct_0/L_x=0.5$. The scattering
angle is $\alpha=\pi/6$.  Temporal evolution of the amplitudes is
traced at the longitudinal positions (a)
$z=V_zct_0/3\approx2.2ct_0$, (c), (d) $z=V_xct_0\approx6.5ct_0$.
The left column shows the SRS evolution inside the glass capillary
($r=0.42$) with the EM seed amplitude at the entrance plane
$a_{s0}=0.4\times10^{-5}$. Plot~(b) shows the saturated solution
for an open-boundary plasma, $r=a_{s0}=0$, for
$z>(V_z/V_x)ct_0\approx3.4ct_0$. In the plot~(d), the long-scale
asymptote~(\ref{19}) of the reflective problem (dashed line) is
compared with the exact solution (solid line)  and with
non-reflective (dotted line) and BSRS (dash-dotted line) solutions
at the point $z=V_zct_0$, $x=L_x/4$.}
\end{figure}

The asymptotic
${}_0\tilde{F}_2(;b_1,b_2;\theta)\sim(2\pi\sqrt3)^{-1}\theta^{(1-b_1-b_2)/3}
e^{3\sqrt[3]{\theta}}+O(\theta^{-1/3})$ at $|\theta|\to\infty$
(Ref.~\onlinecite{IncGamma}) helps to evaluate the level of the EM
seed amplification within the interval $0< z<V_zct_0$,
\begin{equation}
\label{new1}|a_{s+}(z<V_zct_0)|  \sim
a_{s0}\frac{\mathcal{F}}{2\sqrt{3\pi}}\frac{e^{3\sqrt3\zeta^{1/3}/2}}{\zeta^{1/6}},
\end{equation}
where $\zeta\gg1$. Then, $ |N_+|  \sim
(V_z/g_1)(\zeta^{1/3}/z)|a_{s+}|$. At the pulse trailing edge,
$\xi_0=ct_0$, the argument of the asymptotic reaches the maximum
$\zeta_{\max}=(Gct_0/3)^3$ at $z_{\max}=V_zct_0/3$. The EPW
amplitude at this point is
\begin{equation}|N_+(\xi_0,z_{\rm max})| \sim
a_{s0}\frac{G}{g_1}\frac{\sqrt[4]{3}\mathcal{F}}{\sqrt{8\pi}}
\frac{e^{\gamma_0 t_0}}{\sqrt{\gamma_0 t_0}},\label{B12}
\end{equation}
and $|a_{s+}(\xi_0,z_{\rm max})| \sim (g_1/G)|N_+(\xi_0,z_{\rm
max})|$. Equation~(\ref{B12}) shows that the maximum gain on the
transient stage is determined by the 1D temporal
increment~(\ref{13D}) independent on $\alpha$ (the angular
dependence is retained in the pre-exponential factors only). For
the minimally allowed scattering angle,
$\alpha_{\min}=\sqrt{2\omega_{pe}/\omega_0}$ (hence,
$V_z\approx\omega_0/\omega_{pe}$), the point of the maximal gain
is
$z_{\max}(\alpha_{\min})\approx(\omega_0/\omega_{pe})(ct_0/3)\gg
ct_0$. Parameters of the following numerical example
(Fig.~\ref{Fig4}) give $z_{\max}(\alpha_{\min})\approx3$~mm, which
is shorter than a typical capillary length used in experiments
($\ge1$~cm)~\cite{Andreev,Capillary_vacuum,Capillary_plasma,CrosPScr,Andreev_capillary,Kitagawa,CourtoisDiss}.

For $z>z_{\max}$, the entrance effect becomes less pronounced and
finally vanishes at the point $z=V_zct_0=3z_{\max}$, where the EM
signal arriving from $z=-0$ drops behind the laser pulse, and the
instability growth saturates. Evolution of both forward and
backward scattering is then determined by the lateral reflections
only and, given the scattering angle, remains the same in any
$x-y$ cross-section for $z\ge V_zct_0$. We show in
Appendix~\ref{App3} that at $z\ge V_zct_0$, and $\gamma_0t_0>1$, a
cumbersome exact solution~(\ref{11}) tends asymptotically to a
quasi-1D damped mode with a temporal growth rate close
to~(\ref{13D}) that gives the asymptote for the density
perturbation amplitude,
\begin{equation}
N_{+}(x,\xi)  \sim  (N_0/3)[(1-r)/\ln r]e^{s_0\xi-\ln
r(x/L_x)}.\label{19}
\end{equation}
Here, $s_0=(\gamma_0-\Delta\gamma)/c$, $\Delta\gamma/c \approx
-V_x\ln r/(3L_x)$.  Equation~(\ref{19}) is valid under the ``low
leakage'' condition, $r>\exp(-3GL_x/V_x)$, which indicates that
the scattered light is mostly trapped inside a plasma slab: the
energy leakage through the wall at one reflection (that produces
an effective decrement $\Delta\gamma\ll\gamma_0$) is less than the
energy gain on the way between the walls. The lateral growth of
the  asymptote~(\ref{19}) is much slower than the growth with
$\xi$. Asymptotic solution~(\ref{19}) displays the basic result of
the reflective theory of the LA SRS: at large distances from the
entrance plane and large coefficients of amplification, the
amplitudes of unstable waves tend to quasi-1D leaky modes
exponentially growing in time with the increment tending to that
of the BSRS~(\ref{13D}).

Comparison of Eqs.~(\ref{B12}) and~(\ref{19}) shows that the
maximum amplitude of the scattering EPW on the transient stage,
$z<V_zct_0$, differs from the final asymptotic level of density
perturbations roughly by a factor of
$(G/g_1)(\gamma_0t_0)^{-1/2}(a_{s0}/N_0)$. When this factor is
larger than unity, and $|N_{+}(z\ge V_zct_0,x,ct_0)|\sim1$, the
plasma response can become nonlinear in the vicinity of
$z=V_zct_0/3$. This could be avoided by keeping the ratio of the
seed amplitudes $a_{s0}/N_0$ below $(g_1/G)(\gamma_0t_0)^{1/2}$.
The amplitude of the electron density noise is difficult to
control in experiment; however, as shown in Appendix~\ref{App1},
the content of the high-order eigenmodes in the laser radiation
coupled to the capillary (and, hence, the amplitude $a_{s0}$ of
the EM seed signal), can be effectively reduced by increasing the
capillary radius versus the radius of the incident laser beam.

\begin{figure}
\includegraphics[scale=1]{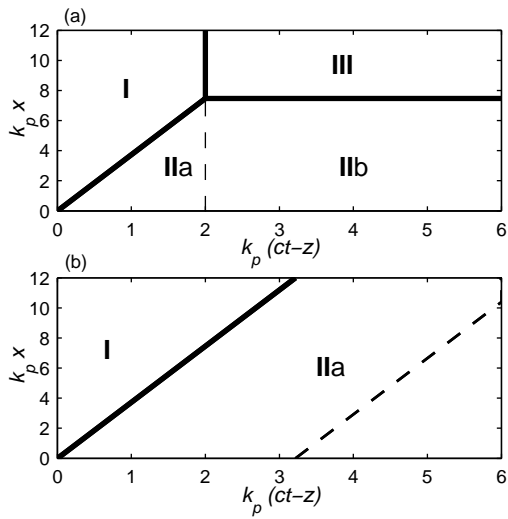}
\caption{\label{Fig5} Ranges of dependence for the SRS under the
angle $\alpha=\pi/6$ in the cross-sections at (a)
$z=V_zct_0/3\approx2.2ct_0$ and (b) $z=V_xct_0\approx6.5ct_0$. The
thick solid line $x=z(V_x/V_z)$ divides the principal ranges
prescribed by the non-reflective theory. The dashed lines are the
characteristics of reflected waves, $x+nL_x=z(V_x/V_z)$,
$\xi=z/V_z$, and $\xi=(x+nL_x)/V_x$. One reflection contributes to
the scattering process in the range {\bf II} in the case (a), and
two reflections
--- in the case (b). Raman amplification of the EM signal given at
the boundary $z=-0$ contributes to the solution in the sub-range
{\bf II}b and range {\bf III}. }
\end{figure}

Figure~\ref{Fig4} shows the spatio-temporal evolution of an
up-going EPW [Eq.~(\ref{12})] for the SRS under the angle
$\alpha=\pi/6$. The laser and plasma parameters are $a_0=0.7$,
$\lambda_0=0.5$~$\mu$m, $ct_0=0.5L_x=6k_p^{-1}$,
$\omega_{pe}/\omega_0=0.007$, which give
$n_0\approx2.2\times10^{17}$~cm$^{-3}$, the pulse duration
$t_0\approx230$~fs, and the maximum increment
$\gamma_0\approx2.25\omega_{pe}$. The level of  EM seed is chosen
$a_{s0}\approx0.58\times10^{-5}a_0$ (according to
Appendix~\ref{App1}, it corresponds to a capillary by a factor of
two wider than in the case of perfect matching; by the definition,
in axi-symmetric geometry, the perfect matching condition provides
coupling 98\% of energy of an incident Gaussian laser pulse to the
fundamental eigenmode EH$_{11}$ of a capillary~\cite{Andreev}).
The level of the plasma noise evaluated in Appendix~\ref{App1} is
$N_0\approx1.5\times10^{-6}$. The plots (a) and (c) correspond to
the plasma confined in a glass capillary with the reflection
coefficient $r=0.42$ (see Fig.~\ref{Fig3}), and (b)  --- to the
unbound plasma ($r=0$ and $a_{s0}=0$); plot (d) shows the
long-term asymptotic behavior of the reflective solution. The
plasma cross-sections are set at (a) $z=V_zct_0/3\approx2.2ct_0$,
(c),~(d) $z=V_zct_0\approx6.5ct_0$. Given the calculation
parameters, the non-reflective solution saturates at
$z=L_x(V_z/V_x)\approx3.4ct_0$ and is exactly the same in any
plasma cross-section $x-y$ beyond that point. Figure~\ref{Fig4}(b)
shows this solution.  The ranges of influence of boundary
conditions are shown in Fig.~\ref{Fig5}. In the range {\bf I} [see
Eq.~(\ref{11B})], the electron density noise from $\xi=0$ is
amplified, and the waves do not experience reflections. In the
range {\bf II}a the instability is seeded by the free-plasma
noise, and yet is enhanced by the reflections; the Raman amplified
signal arriving from the entrance plane is added to these waves in
the range {\bf II}b. The EM seed from the entrance plane $z=-0$ is
amplified by the forward SRS in the range {\bf III.}

In a capillary, the forward SRS is considerably enhanced at
$z<V_zct_0$ (roughly by a factor of $70$ in amplitude) versus the
case of unbound plasma [compare Figs.~\ref{Fig4}(a) and~(b)].
Despite the EM seed amplified in the range {\bf II}b is
non-exponentially growing, the high ratio of seed amplitudes,
$a_{s0}/N_0\approx2.7\gg(g_1/G)(\gamma_0t_0)^{1/2}\approx0.026$,
makes the entrance effect rather pronounced [plot~\ref{Fig4}(a)].
For $N_0\approx1.5\times10^{-6}$, and $\mathcal{F}\sim 1$,
Eq.~(\ref{B12}) gives $\log_{10}|N_+(\xi_0,z_{\rm max})/N_0|
\approx 7.15$, which agrees with Fig.~\ref{Fig4}(a). Hence,
parameters of the numerical example lay at the border of validity
of the linear approach, and increase in $a_{s0}$ will result in
the nonlinearity of the plasma response on the transient stage
\{e.g., perfect matching gives $a_{0s}\approx0.012a_0$, hence,
according to Eq.~(\ref{B12}), $|N_+(\xi_0,z_{\rm max})| \sim
10^{6}$; this burst of the forward SRS has been the regular
feature in our numerical experiments and in the fluid
simulations~\cite{AndreevPrComm}\}. On the other hand, reduction
$a_{s0}$ to the level $7\times10^{-8}a_0$ (which in the
axi-symmetric case would correspond to a capillary tube by a
factor 2.5 wider than in the case of perfect matching, see
Appendix~\ref{App1}), will make the forward Raman amplification of
the EM seed almost negligible and thus tolerable on the transient
stage.

Plot~\ref{Fig4}(c) shows a quasi-1D saturated solution [compare
with Fig.~\ref{Fig4}(b)] thus shaped by the contribution from two
reflections [according to Fig.~\ref{Fig5}(b)], which, in full
agreement with the long-scale asymptote~(\ref{19}), demonstrates
the growth rate close to that of BSRS. Difference between
Figs.~\ref{Fig4}(a) and~(c) shows that, under the parameters of
our example, the forward scattering ($\alpha<\pi/2$) is
characterized at $z<V_zct_0$ by much higher gain than the SRS in
the backward direction ($\alpha>\pi/2$). This situation is
completely reverse of the SRS in an unbounded plasma, where the
gain can only fall as an angle drops~\cite{Mora0}. So, the higher
the EM seed level produced by the laser beam coupling (that is,
the tighter the capillary in a numerical or real-scale
experiment), the more important becomes the forward SRS. In such
case, a high amplification level of waves may be observed within
quite a long distance in plasma,
$z<V_z(\alpha_{\min})ct_0\approx(\omega_0/\omega_{pe})ct_0$ [see
also the discussion following Eq.~(\ref{B12})]. This effect is
adverse for such applications as the self-modulated LWFA in
capillaries~\cite{CourtoisDiss}. The way of reducing the excessive
forward SRS enhancement can be found in using a wider capillary
(the laser focal spot fixed) than the perfect matching requires.

Figure~\ref{Fig4}(d) traces the temporal evolution of the up-going
EPW at $x=L_x/4$ (near the capillary wall). The
asymptote~(\ref{19}) perfectly approximates (and, for
applications, can  be used instead of) the exact reflective
solution~(\ref{12}). The dash-dotted and dotted lines in
Fig.~\ref{Fig4}(d) correspond  to the BSRS solution
$|N_+(\xi)|\approx(N_0/3)\exp \left(\gamma_0\xi/c\right)$ and the
exact non-reflective solution, respectively, providing the upper
and lower limits of the convective gain variation. In this
example, only in the vicinity of the border $x=0$ the coefficients
of amplification in the reflective and non-reflective cases are
considerably different [compare Figs.~\ref{Fig4}(b) and~(c)].
Contribution from the reflections increases the wave amplitude at
$x=L_x/4$ by roughly an order of magnitude (compare solid and
dotted lines at $k_p\xi\approx6$). For the parameters chosen, the
scattering EPW remains linear in the convective saturated regime
($z>V_zct_0$): the plasma noise level,
$N_0\approx1.5\times10^{-6}$, substituted into Eq.~(\ref{19}),
gives $|N_+|\le0.2$ throughout the whole time interval $0<t<t_0$.

In the limit $r\to1$ the total suppression of the lateral
convection occurs. The up- and down-going amplitudes~(\ref{C})
become purely one-dimensional for $z>V_zct_0$ and completely
identical. These amplitudes grow in time exponentially with the
BSRS increment~(\ref{13D}).

\section{\label{Sec4}Conclusion}

We have proposed a 2D non-stationary linear theory of strongly
coupled LA SRS of a short laser pulse in a flat plasma slab
confined between mirror-reflecting walls (flat capillary). In a
capillary, the lateral convection of scattered light is partly
suppressed by the oblique reflections, and the instability
experiences an enhancement. Additional enhancement of the SRS in
forward direction ($\alpha>\pi/2$) is produced by the
amplification of the electromagnetic seed signal that is formed of
the high-order capillary eigenmodes at the entrance plane
(formation of the signal is a consequence of the laser beam
coupling to the capillary). The convective nature of LA SRS does
not change. The asymptotic behavior of the waves demonstrates the
transition from the set of 2D modes to the dominant quasi-1D
damped mode. Even for near-forward scattering the convective gain
of the dominant quasi-1D mode may be close to the BSRS gain.

\begin{acknowledgments}
This work was started with the aid of a postdoctoral fellowship
from the Centre de Physique Th\'eorique, Ecole Polytechnique, and
its completion was supported by the Fortbildungsstipendium from
Max-Planck-Institut f\"ur Quantenoptik.  We acknowledge useful
discussions with N. E. Andreev, B. Cros, L. M. Gorbunov, G.
Matthieussent, and J. Meyer-ter-Vehn.
\end{acknowledgments}

\appendix

\section{\label{App1} Seed sources for LA SRS}
The LA SRS under arbitrary angle in a strongly rarefied plasmas
($\omega_{pe}\ll\omega_0$) is seeded by spontaneous electron
density fluctuations ahead of the pulse. A root-mean-square (rms)
amplitude of these fluctuations is represented in the equations by
the quantity $N_0$, which gives the amount of seed corresponding
to the element of solid angle $d\Omega_{{\bf k}_e}$ in the
direction of the wave vector ${\bf k}_e$ of scattering EPW, and
can be expressed as $ N_0^2(k_e)=d\Omega_{{\bf k}_e}\int
n_e^{3D}k^2\,dk\approx n_e^{3D}(k_e)k_e^2\Delta k_e d\Omega_{{\bf
k}_e}$, where $n_e^{3D}(k_e)$ is a spectral density of electron
fluctuations integrated over frequencies~\cite{Akhiezer}. The
integral is taken over the area of maximal spectral density of
scattering EPW [$k\approx k_e$, the amplification bandwidth
$c\Delta k_e \approx 4\sqrt{\gamma_0/t_0}$  is estimated at
$\gamma_0t_0\gg1$ with taking account of the gain narrowing
(Ref.~\onlinecite{Kalmykov_d})]. Using the ratio of the phase
volumes $|d{\bf k}_e /d{\bf k}_s|=2\sin^2(\alpha/2)$
(Ref.~\onlinecite{Sakharov2000}), we express the seed amplitude
through the element of solid angle $d\Omega_{{\bf k}_s}$ in the
direction of detector, $N_0^2\approx 8n_e^{3D}(k_e)k_0^2
[(\gamma_0/c)/(ct_0)]^{1/2} \sin^2(\alpha/2)d\Omega_{{\bf k}_s}$.
Our theoretical formalism based on the assumption of quasi-plane
interacting waves requires small variation of the scattered wave
amplitude across the direction ${\bf k}_s$ in the transversely
limited area, $0\le x \le L_x$, which gives an estimate of the
angular spread $\Delta \alpha\approx\cos\alpha/(L_xk_0)$. The
element of solid angle then estimated as $\Delta\Omega_{{\bf
k}_s}\approx2\pi\sin\alpha\Delta\alpha=\pi\sin2\alpha/(L_x k_0)$
gives
\begin{equation}
\label{A1} |N_0| \! \approx \!
\sqrt{\frac{8\pi}{n_0\lambda_0^3}\frac{1 \! + \! (k_e{r_D}_e)^2}{2
\! + \!
(k_e{r_D}_e)^2}\frac{(\gamma_0t_0)^{1/2}}{(\omega_0t_0)^{3/2}}\frac{\sin^2(\alpha/2)
\sin2\alpha}{k_0L_x}},
\end{equation}
where the spectral density of low-frequency electron fluctuations
$n_e^{3D}(k_e)$ is evaluated using formula~(11.2.6.6) of
Ref.~\onlinecite{Akhiezer}. Parameters of Fig.~\ref{Fig4} and
$k_e{r_D}_e\ll1$  give $|N_0|\approx1.5\times10^{-6}$.

\begin{figure}
\includegraphics[scale=1]{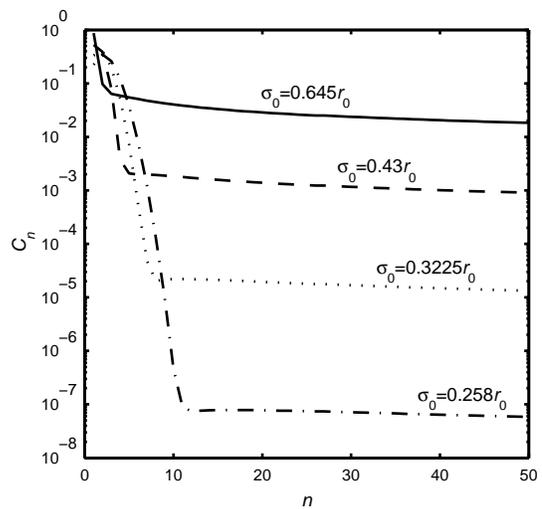}
\caption{\label{Fig6} Overlap integral between the radial profiles
of amplitude of the incident laser (Gaussian) and of the capillary
hybrid eigenmodes EH$_{1n}$. }
\end{figure}

For the strongly coupled SRS in the forward direction
($\alpha<\pi/2$), coupling the laser beam to a capillary creates
additional source of instability.  The radial profile of an
incident radiation with the wings cut off by the edges of the
entrance aperture is approximated with an expansion through an
infinite number of radial eigenmodes (having the same frequency
$\omega_0$)~\cite{Marcatili,Andreev}. The high-order eigenmodes
(characterized by the frequency $\omega_s=\omega_0$ and high
transverse wavenumbers, $k_{n\perp}\sim {k_{s}}_x$, where the
integer $n$ is the mode order) form the seed signal that is
further amplified in plasma by the SRS [in the model form, the
transverse profile of this signal is given by Eq.~(\ref{B21})]. It
should be emphasized that these modes provide no seed for the
weakly coupled SRS (including near-forward SRS and RMI~\cite{RMI}
that correspond to small scattering angles,
$\alpha\ll\sqrt{\omega_{pe}/\omega_0}$), as this process requires
the frequency matching $\omega_s\approx\omega_0-\omega_{pe}$
between the seed and the pump.

Exact functional form of the capillary eigenmodes depends on the
geometry chosen. Despite the LA SRS in a flat capillary is
considered in the paper, we suppose that an estimate of the EM
seed level will be more useful for applications if inferred from
the axi-symmetric theory of the laser beam propagation in a
dielectric tube~\cite{Marcatili,Andreev}. The theory represents an
electric field profile at the entrance aperture of the radius
$r_0$ as an infinite sum $\tilde{a}(r)=a_0\sum_{n=1}^{+\infty}
C_nJ_0(k_{n\perp }r)$~(Ref.~\onlinecite{CrosPScr}), where
$C_n=2[r_0J_1(u_n)]^{-2}\int_0^{r_0}a(r)J_0(u_n r/r_0)r\,dr$ is
the overlap integral of the hybrid capillary eigenmode  EH$_{1n}$
with the incident laser profile $a(r)=\exp(-r^2/\sigma_0^2)$
(Fig.~\ref{Fig6}). Here, $u_n$ is the $n$th zero of the zero-order
Bessel function of the first kind, $J_0(u_n)=0$; for $k_0r_0\gg1$,
and $n\gg1$, $k_{n\perp}r_0\approx u_n\approx (n+1/2)\pi$.
Fig.~\ref{Fig6} shows that about 98\% of laser energy is coupled
to the fundamental mode for $\sigma_0=0.645r_0$ (the perfect
matching condition); however, the overlap integral decays very
slowly as $n$ grows \{the analytic fit, $C_n\sim
F_n=\exp[-(n+50)^{0.3}]$, is almost exact for $n>10$\}. When the
ratio $\sigma_0/r_0$ drops, the larger number of lower-order
eigenmodes is effectively excited (up to 5 for
$\sigma_0=0.258r_0$), but contribution from the higher-order modes
into the radiation profile at $z=-0$ drops sharply (e.g., analytic
fit $C_n\approx3\times10^{-6}F_n$, $n>10$, holds for
$\sigma_0=0.258r_0$). Therefore, choosing wider capillary is the
way of reducing the effect of laser coupling on the forward
strongly coupled SRS.

The numerical example of subsection~\ref{SubSec3.2} shows how the
SRS under the angle $\alpha=\pi/6$ develops in the capillary with
an EM seed amplitude characteristic of the axi-symmetric capillary
tube by a factor of two wider than in the case of perfect
matching. For the parameters of Fig.~\ref{Fig4}, with the value
$L_x/2\approx860k_0^{-1}$ assigned to $r_0$, equalizing the
effective radial wave number $k_{n\perp}\approx n\pi/r_0$ to the
resonant wave number $k_{s\perp}=k_0\sin\alpha$ gives the resonant
mode order $n^*\approx(k_0r_0/\pi)\sin\alpha\approx136$
corresponding to $C_{136}\approx0.58\times10^{-5}$ under the
condition $\sigma_0=0.3225r_0$ (analytic fit
$C_n\approx0.7\times10^{-3}F_n$ was used; see also
Fig.~\ref{Fig6}). One can expect that several modes with $n\approx
n^*$ can contribute to the effective amplitude $a_{s0}$ of the
seed signal. The difference in the mode numbers $\Delta n=n^*-n$
causes the angular spread $\Delta\alpha$ around the given
scattering angle; this spread should not exceed the admissible
value established above, $\Delta
\alpha<\cos\alpha/(L_xk_0)\approx\cos\alpha/(2r_0k_0)$. For
$|\Delta\alpha|\ll\alpha$ and $|\Delta n|\ll n^*$, one has $\Delta
n\approx(k_0 r_0/\pi)\cos\alpha
\Delta\alpha<\cos^2\alpha/(2\pi)<1$, so that only one capillary
mode with $n=n^*$ can contribute to the scattering under the given
angle. Therefore, the seed amplitude used for the numerical
demonstration of Fig.~\ref{Fig4}(a) is evaluated as $a_{s0}\approx
C_{136}a_0\approx 0.4\times10^{-5}$.

\section{\label{App2} Derivation of the exact reflective solution}

Omitted $k_p^2$ in the LHS of Eq.~(\ref{4}), the Laplace transform
(LT) of Eqs.~(\ref{A}) with respect to $\xi$ (LT variable $s$) and
with respect to $z$ (LT variable $p$) gives the set of ODE for the
Laplace images $\bar{\bar{a}}_{s\pm}(x,p,s;r)$,
\begin{equation}
\left(\frac{\partial}{\partial x} \mp\Omega\right)
\bar{\bar{a}}_{s\pm} = \pm K(x), \label{A_3}
\end{equation}
where $\Omega=(V_z/V_x)(\Gamma_s-p)$,
$\Gamma_s=(i\,G^3/s^2-s)/V_z$,
$K(x)=K_1/(sp)+(K_2/s)[1-(2x/L_x-1)^2]$, $K_1=g_1N_0/(iV_x)$,
$K_2=(V_z/V_x)a_{s0}$. The boundary conditions are $ {\bar{\bar
a}}_{s+}(0,p,s) = r{\bar{\bar a}}_{s-} (0,p,s)$ and $ {\bar{\bar
a}}_{s-}(L_x,p,s) =r {\bar{\bar a}}_{s+}(L_x,p,s)$.
Equations~({\ref{A_3}) admit the solution \begin{widetext}
\begin{eqnarray}
\nonumber\bar{\bar a}_{s+}(x,p,s;r) & = & -
\left(\frac{K_1}{p}+K_2\right)\left[1 - \frac{(1-r)e^{\Omega
x}}{1-re^{\Omega L_x}}\right]\frac{1}{s\Omega} +
\frac{K_2}{(L_x/2)^2}\left\{1 + \left[\Omega\left(\frac{L_x}{2} -
x \right)-1\right]^2\right\}
\frac{1}{s\Omega^3}\nonumber\\
& - &  \frac{K_2}{(L_x/2)^2} \frac{e^{\Omega x}}{1-re^{\Omega
L_x}} \left\{1 + \left(\Omega \frac{L_x}{2} - 1\right)^2-
r\left[1+\left(\Omega \frac{
L_x}{2}+1\right)^2\right]\right\}\frac{1}{s\Omega^3}, \label{A_7}
\end{eqnarray}
where $x_n\equiv x+nL_x$. Lateral symmetry gives ${\bar{\bar
a}}_{s-}(x)  = {\bar{\bar a}}_{s+}(L_x-x)$.  The Laplace transform
inversion of Eq.~(\ref{A_7})  with respect to $p$ reads
\begin{eqnarray}
\nonumber{\bar a}_{s+}(x,z,s;r) & = & -
\frac{\tilde{K}_1}{s\Gamma_s} \! + \!
\left[\tilde{K}_1\frac{e^{\Gamma_s z}}{s\Gamma_s} \! - \!
\tilde{K}_2\frac{e^{\Gamma_s z}}{s}\frac{(x \! - \! \tilde{z})(x
\! - \! \tilde{z} \! - \! L_x)}{(L_x/2)^2}\right]H(x \! - \!
\tilde{z}) \! + \! \tilde{K}_1 (1 \! - \! r)\sum_{n=0}^{\infty}
r^n\frac{e^{\Gamma_s(V_z/V_x)x_n}}{s\Gamma_s}H(\tilde{z} \! - \! x_n)\\
& + &   \sum_{n=1}^{\infty} r^n\left\{
\tilde{K}_1\frac{e^{\Gamma_sz}}{s\Gamma_s} -
\tilde{K}_2\frac{e^{\Gamma_sz}}{s}\frac{(x_{n-1}-\tilde{z})(x_n-\tilde{z})}{(L_x/2)^2}\right\}
\left[H(\tilde{z}-x_{n-1})-H(\tilde{z}-x_n)\right]\label{A_9},
\end{eqnarray}
\end{widetext}
where, in accordance with Ref.~\onlinecite{McKinstrie3}, the
expansion $[1-r\exp(\Omega L_x)]^{-1}=\sum_{n=0}^\infty r^n
\exp(n\Omega L_x)$ is used, and $\tilde{z}\equiv(V_x/V_z)z$,
$\tilde{K}_{1,2}\equiv(V_x/V_z)K_{1,2}$. Equation~(\ref{A_9})
includes Laplace images of the three types: $1/(s\Gamma_s)$,
$e^{\Gamma_sy}/(s\Gamma_s)$, and $e^{\Gamma_sy}/s$. Their
inversions read:
\begin{equation}
\label{A_12} {\cal{L}}_s^{-1}\left\{e^{\Gamma_s
y}/s\right\}=\sqrt{\pi}H(\xi-z/V_z)
{}_0\tilde{F}_2(;1,1/2;i\zeta),
\end{equation}
expressed through the regularized generalized hypergeometric
function~\cite{IncGamma} of variable
$i\zeta=i(G^3/4)(\xi-z/V_z)^2z/V_z$;
\begin{equation}
{\cal{L}}_s^{-1}\left\{\frac{1}{s\Gamma_s}\right\}
 =
-\frac{V_z}{3}\sum_{j=1}^{3}\frac{e^{c_j\xi}}{c_j},\label{A_10}
\end{equation}
where $c_j^3=iG^3$, so that $c_1=-iG$, $c_{2}=(i+\sqrt3)G/2$,
$c_{3}=(i-\sqrt3)G/2$; and
\begin{equation}
\label{A_14}{\cal{L}}_s^{-1}\left\{\frac{e^{\Gamma_s
y}}{s\Gamma_s}\right\}=-\frac{V_z}{3}
\sum_{j=1}^3\frac{e^{c_j\xi}}{c_j}F_{s}\left(c_j,\frac{y}{V_z},\xi\right),
\end{equation}
which is expressed through the fundamental
solutions~\cite{Kalmykov}
\begin{equation}
\label{A_11} F_s(\mu,v,\xi)  =  e^{-\mu
v}\sum\limits_{n=0}^{\infty} \frac{(\mu
v)^n}{n!(2n)!}\gamma\bigl(2n+1,\mu(\xi-v)\bigr)H(\xi-v),
\end{equation}
where $\gamma(m,\tau)=\int_0^\tau e^{-\tau'}{\tau'}^{m-1}\,d\tau'$
is the incomplete gamma-function of the order $m$ of a complex
variable $\tau$ (Ref.~\onlinecite{IncGamma}). Combining
expressions~(\ref{A_10}),~(\ref{A_14}), and~(\ref{A_12}) in
Eq.~(\ref{A_9}) gives the envelope~(\ref{11}) of the up-going EMW.
The amplitude~(\ref{12}) of the scattering EPW is derived in the
similar fashion.

\section{\label{App3}  Asymptotic reflective solution}

The asymptotic can be found by applying the inversion formula, $
{\bar a}_{s+}(x,z,s)=(2\pi
i)^{-1}\int_{c-i\infty}^{c+i\infty}e^{px} {\bar{\bar
a}}_{s+}(x,p,s)\,dp$, to the expression~(\ref{A_7}). The
asymptotic behavior at $z\to\infty$ is determined by the
singularities of the integrand at $p=0$, $p=\Gamma_s$, and
$p_n=\Gamma_s-\nu+(V_x/V_z)(2\pi n i/L_x)$ with $n$ integer and
$\nu=-(V_x/V_z)(\ln r/L_x)$.  Expanding ${\bar{\bar
A}}_{s+}(x,p,s)$ in the vicinity of the specific points shows that
all the singularities at $p=\Gamma_s$ and $p=p_n$ give the
contribution $\bar{a}_1(x,z,s)\propto\exp(\Gamma_sz)$. Therefore,
${\cal{L}}_s^{-1}\{\bar{a}_1\}\propto H(\xi-z/V_z)$, and, for a
pulse of finite duration, $\xi\le ct_0$, no contribution comes to
the asymptotic from these specific points as soon as the distance
$z$ from the entrance plane exceeds $V_zct_0$ [or $0$, for
$\alpha>\pi/2$]. (Actually, the Laplace image singularities at
$p=\Gamma_s$ and $p=p_n$ determine the entrance effect, i.e., the
waves produced by the seed at the entrance aperture amplified in
the pump field in plasma; these waves will inevitably drop behind
the laser pulse and their effect therefore vanishes at
$z\to\infty$.) The contribution from $p=0$,
\[
{\bar a}_{s+}(x,z\to\infty,s)  \sim \frac{1}{s\Gamma_s}
\left\{1-\frac{(1-r)\exp[(V_z/V_x)\Gamma_s x
]}{1-r\exp[(V_z/V_x)\Gamma_s L_x]}\right\}\label{A_15_2},
\]
thus determines the long-scale evolution of the instability in
plasma, which is dominated by the lateral reflections of scattered
EMW. Neither  $s=c_j$ (where $\Gamma_s=0$) nor $s=0$ are the
singularities of the image ${\bar a}_{s+}(x,z\to\infty,s)$, so
contributions to the asymptotic originate from the singular points
$s_n$ only, which are the solutions of the equation
$\Gamma_{s_n}=\nu_n$ [here, $\nu_n=(V_x/V_z)(-\ln r+2\pi i
n)/L_x$], or $s_n^3+\nu_n V_zs_n^2-iG^3=0$, $n$ is integer. The
fundamental specific point $s_0$ with the maximum real part is the
root of the cubic equation $s^3+\nu_0V_z s^2-iG^3=0$ which, for
arbitrary $\nu_0$, admits quite a cumbersome explicit expression.
We address to the physically interesting limit of ``low leakage'',
i.e., $\nu_0 <3G/V_z$, or $r>\exp(-3GL_x/V_x)$, which means that
the scattered EMW gains more energy between two reflections than
loses due to leakage through the wall at one reflection. In this
case the lateral convection is mostly suppressed. Solution
obtained via the perturbation approach reads
$s_0\approx(i+\sqrt3)G/2-\nu_0 V_z/3$. Contribution to the
asymptotic from that point is of the order of $e^{s_0\xi}$, which
grows in time with an increment ${\rm Re}\,s_0$. However, the
absolute value of $\nu_n$ grows with $n$, so an evaluation has to
be done of the contribution from the points $s_n$ with large $n$.
We again use the perturbation approach with the small parameter
$\mu_n=G/(|{\tilde\nu}_n|V_z)$, that is, $n>GL_x/(2\pi V_x)$, and
find the solutions $s_n^{(1)}\approx-\nu_nV_z-iG\mu_n^2$,
$s_n^{(2,3)}\approx \pm G\sqrt{-({\rm sign} \,n)\mu_n}$.
Obviously, ${\rm Re}s_j\ll G$ for $\mu_n\ll 1$, so that
contribution from these points to the asymptotic is negligible
compared to that from the points $s_n$ with $n<GL_x/(2\pi V_x)$,
which is of the order of $e^{s_0\xi}$. Expanding the image ${\bar
a}_{s+}(x,z\to\infty,s)$ in the vicinity of $s=s_0$, we arrive at
the expression asymptotically valid for $G\xi\gg1$,
\[ a_{s+}(x,\xi)  \sim  \frac{V_z}{G}\left(\frac{1
 -  r}{\ln r}\right)\frac{i  -  \sqrt3}{6}\exp\left(s_0\xi
-  \ln r\frac{x}{L_x}\right),
\]
which represents the unstable EMW as a quasi-1D exponentially
growing mode.


\end{document}